\documentclass[preprint,12pt,authoryear]{elsarticle}

\usepackage{amssymb}
\usepackage{amsmath}
\usepackage{amsmath,amssymb,amsfonts}
\usepackage{caption}
\usepackage[figuresright]{rotating}
\usepackage{algorithmic}
\usepackage{algorithm}
\usepackage{graphicx}
\usepackage{svg}
\usepackage{algorithmic}
\usepackage{algorithm}
\usepackage{graphicx}
\usepackage{textcomp}
\usepackage{xcolor}
\usepackage{multirow}
\usepackage{CJKutf8}
\usepackage{float}
\usepackage{threeparttable}
\usepackage{verbatim}
\RequirePackage{pdflscape}
\usepackage{textcomp}
\usepackage{xcolor}
\usepackage{array}    
\usepackage{makecell} 
\usepackage{multirow}
\usepackage{float}
\usepackage{eurosym}
\usepackage{booktabs}
\usepackage{subfig}
\usepackage{svg}
\usepackage{color}
\usepackage[utf8]{inputenc}
\usepackage{enumitem}
\usepackage{hyperref}
\hypersetup{
    colorlinks=true,
    linkcolor=black,
    citecolor=black, 
    urlcolor=black
}

\makeatletter
\def\ps@pprintTitle{%
 \let\@oddhead\@empty
 \let\@evenhead\@empty
 \let\@oddfoot\@empty
 \let\@evenfoot\@oddfoot
}
\makeatother

\journal{Journal of Cleaner Production}

\begin{document}

\begin{frontmatter}



\title{Carbon Disclosure Effect, Corporate Fundamentals, and Net-zero Emission Target: Evidence from China} 

\author[a]{Xiyuan Zhou}
\author[b]{Xinlei Wang}
\author[c]{Xiang Fei}
\author[a]{Wenxuan Liu}
\author[d]{Bai-Chen Xie}
\author[e]{Junhua Zhao}

\affiliation[a]{organization={School of Electrical and Electronic Engineering, Nanyang Technological University},
            city={Singapore},
            postcode={639798}, 
            country={Singapore}}

\affiliation[b]{organization={School of Electrical and Information Engineering, The University of Sydney},
    city={Sydney},
    postcode={NSW2006}, 
    country={Australia}}

\affiliation[c]{organization={School of Data Science, The Chinese University of Hong Kong, Shenzhen},
            city={Shenzhen},
            postcode={518100}, 
            country={China}}

\affiliation[d]{organization={College of Management and Economics,Tianjin University},
            city={Tianjin},
            postcode={300072}, 
            country={China}}
            
\affiliation[e]{organization={School of Science and Engineering, The Chinese University of Hong Kong, Shenzhen},
            city={Shenzhen},
            postcode={518100}, 
            country={China}}

\tnotetext[1]{X. Zhou and X. Wang contributed equally to this work and should be regarded as co-first authors.}

\tnotetext[2]{Correspondence should be addressed to Junhua Zhao at the School of Science and Engineering, The Chinese University of Hong Kong, Shenzhen, Guangdong 518100, China (email: zhaojunhua@cuhk.edu.cn), and Bai-Chen Xie, College of Management and Economics, Tianjin University, Tianjin 300072, China (email: xiebaichen@126.com).}

\begin{abstract}
In response to China's national carbon neutrality goals, this study examines how corporate carbon emissions disclosure affects the financial performance of Chinese A-share listed companies. Leveraging artificial intelligence tools, including natural language processing, we analyzed emissions disclosures for 4,336 companies from 2017 to 2022. The research demonstrates that high-quality carbon disclosure positively impacts financial performance with higher stock returns, improved return on equity, increased Tobin's Q ratio, and reduced stock price volatility. Our findings underscore the emerging importance of carbon transparency in financial markets, highlighting how environmental reporting can serve as a strategic mechanism to create corporate value and adapt to climate change.
\end{abstract}





\end{frontmatter}




\section{Introduction}
In response to global climate challenges, China announced its dual-carbon target in 2020 and established a national carbon market in 2021, creating a market-based mechanism for emissions regulation \citep{liu2022quality,zhou2024carbon}. This policy shift has transformed carbon disclosure into a critical tool for stakeholder communication and risk management. Our study examines how such disclosure impacts the financial performance of Chinese-listed companies in this evolving regulatory landscape.

Growing awareness of climate risks has driven global regulatory requirements for carbon disclosure, exemplified by the Stock Exchange of Hong Kong (HKEX) and US Securities and Exchange Commission (SEC) proposals in 2022 mandating comprehensive scope 1, 2, and 3 emissions reporting \citep{alsaifi2020carbon,liu2024comprehensive,securities2022sec}. As carbon disclosure becomes a normalized practice globally, investors increasingly incorporate this information into their valuation models, making the relationship between carbon disclosure and financial performance increasingly critical.

The growing importance of carbon disclosure necessitates a deeper understanding of its financial implications. While enhanced carbon disclosure can improve resource allocation and corporate image through reduced information asymmetry and climate risk management, it may also introduce costs and potential market penalties for negative environmental performance  \citep{bae2018cross,sullivan2012does,endrikat2016market,lee2015impacts}. This raises questions about the financial implications of carbon disclosure. Meanwhile, China's unique institutional environment, characterized by intertwined regulatory frameworks and coexisting regional-national carbon markets, provides a unique setting to examine how carbon disclosure affects firm performance under complex regulatory conditions \citep{siddique2021carbon, downar2021impact}.

Existing studies on carbon disclosure and financial performance face three major limitations. First, previous research typically examines isolated financial metrics, such as focusing solely on stock returns or accounting performance \citep{dixon2013beyond,endrikat2014making}. This overlooks the multifaceted nature of financial performance and fails to capture how carbon disclosure simultaneously affects market valuation, risk perception, and operational efficiency. The absence of a comprehensive analysis framework limits our understanding of the various channels through which environmental transparency influences firm value.

In addition, previous studies examine the independent factors that influence disclosure decisions without considering their interdependencies \citep{alsaifi2020carbon}. For example, while research has separately examined the roles of ownership structure, market competition, and regulatory pressure, the dynamic interactions among these factors remain poorly understood. This limitation is particularly pronounced in complex institutional environments where multiple forces simultaneously shape firms' disclosure strategies.

Moreover, empirical investigation of these relationships is hampered by data constraints, as existing carbon disclosure databases focus mainly on national or regional aggregates, with limited granular insight at the company level \citep{khan2016corporate}. The Carbon Disclosure Project (CDP), while pioneering, suffers from low coverage of Chinese companies and relies primarily on questionnaire-based data collection. These limitations are particularly salient in China's context, where the dual-carbon target has fundamentally altered the cost-benefit dynamics of environmental disclosure, creating an urgent need for sophisticated, technology-enabled approaches to analyze corporate carbon disclosure strategies comprehensively.

Based on these research gaps, we make three main contributions to the literature:
\begin{enumerate}[label=(\arabic*)]
\item We address data limitation by building a comprehensive carbon disclosure finance dataset that covers 4,336 listed Chinese A-share companies from 2017 to 2022. Our dataset leverages artificial intelligence technologies to systematically extract and analyze carbon disclosure information, enabling unprecedented insights into corporate environmental transparency.

\item We provide novel empirical evidence on the heterogeneous effects of carbon disclosure through a multi-dimensional analysis framework. Our approach integrates both market-based (stock returns, price volatility) and accounting-based (ROE, Tobin's Q) measures to comprehensively capture how carbon disclosure affects firm financial performance. This analysis reveals significant variations in financial impact between high-carbon and low-carbon industries following the dual-carbon target implementation.

\item We advance the understanding of carbon disclosure determinants by developing an integrated framework that captures the dynamic interactions between firm-level characteristics and institutional factors. Our analysis reveals how internal factors (R\&D investment, overseas listing status) and external conditions (carbon market participation, state ownership, industry type) jointly shape firms' disclosure strategies in China's unique regulatory environment.
\end{enumerate}

The remainder of this paper proceeds as follows. Section \ref{sec:Relevant Literature Review and Hypothesis} reviews literature and presents hypothesis development. Section \ref{sec:Institutional background, database, variables, and sample} describes the data. Section \ref{sec: Empirical results}, we present the main empirical results and interpretation. Section \ref{sec:Conclusion and policy implications} concludes and proposes policy implications.

\section{Literature Review and Hypothesis}\label{sec:Relevant Literature Review and Hypothesis}

\subsection{Carbon Disclosure and Company Performance}

The relationship between carbon disclosure and financial performance operates through multiple channels.Carbon disclosure can affect firm value through both direct and indirect mechanisms, with the net effect dependent on the institutional context and market environment. On the positive side, carbon disclosure can enhance firm value through several channels. First, improved information transparency reduces information asymmetry between firms and investors, leading to more efficient resource allocation in capital markets \citep{bae2018cross}. Second, proactive carbon disclosure signals strong environmental management capabilities, potentially enhancing corporate reputation and stakeholder trust \citep{luo2014does}. Third, transparent environmental reporting can reduce climate-related risks and associated costs \citep{sullivan2012does}. However, carbon disclosure also introduces potential costs and risks. Companies face direct costs in collecting and verifying emissions data, as well as potential market penalties for negative environmental performance \citep{endrikat2016market,lee2015impacts}. Additionally, disclosure may require increased environmental investments that could strain financial resources \citep{reverte2009determinants,eccles2014impact}.

In China's context, we argue that the benefits of carbon disclosure will outweigh the costs following the dual-carbon target implementation for several reasons. First, the policy signals a long-term commitment to emissions reduction, making environmental transparency increasingly valuable for stakeholder relationships. Second, early adopters of comprehensive disclosure practices may gain competitive advantages in accessing green financing and government support. Third, the establishment of national and regional carbon markets creates additional incentives for transparent emissions reporting. Therefore, we hypothesize:

\textbf{H1. } Following the implementation of the dual-carbon target, carbon disclosure positively impacts corporate financial performance.

\subsection{Determinants of Carbon Disclosure}

Understanding the determinants of carbon disclosure is critical for investigating its financial implications. Prior literature has identified several theoretical frameworks explaining firms' disclosure decisions: legitimacy theory, stakeholder theory and economics-based theories.

First, legitimacy theory suggests that carbon disclosure emerges as a response to external pressures, particularly regulatory requirements and social expectations \citep{patten2002relation,tang2020external}. In China's context, while environmental disclosure regulations primarily target key polluting firms, the dual-carbon target and carbon market establishment have created broader institutional pressures for transparency \citep{li2019motivations}. Secondly, stakeholder theory frames carbon disclosure as a strategic response to diverse stakeholder demands \citep{roberts1992determinants,luo2012corporate}. Firms disclose carbon information to maintain stakeholder relationships, particularly in response to public and governmental concerns. This is especially relevant in China's state-dominated economy, where firms must balance multiple stakeholder interests. Thirdly, economics-based theories emphasize the cost-benefit analysis underlying voluntary disclosure decisions \citep{verrecchia2001essays,clarkson2008revisiting}. Firms may disclose to reduce information asymmetry and financing costs, particularly when they possess favorable environmental performance \citep{healy2001information,luo2014does}.

Drawing on these theoretical perspectives and China's unique institutional setting, we posit that both internal characteristics (R\&D investment, overseas listing) and external factors (carbon market participation, state ownership, industry type) shape firms' disclosure decisions. Therefore, we hypothesize:

\textbf{H2.} A firm's propensity for carbon disclosure is positively associated with its R\&D investment, overseas listing status, carbon market participation, state ownership, high-carbon industry affiliation, and implementing the dual-carbon target.

\subsection{Carbon Disclosure Measurement}

The measurement of corporate carbon disclosure has evolved through two main methodologies: questionnaire-based approaches and content analysis. The Carbon Disclosure Project (CDP), while pioneering in collecting company-level emissions data, faces significant limitations, particularly in emerging markets. For instance, only about 100 Chinese companies participate in CDP reporting, primarily large corporations, resulting in limited coverage across industries and firm sizes \citep{luo2014does}.

Content analysis through keyword searching offers an alternative approach, examining terms such as carbon disclosure, carbon reporting, and greenhouse gas emissions disclosure in public reports \citep{borghei2021carbon}. With the rapid development of artificial intelligence and machine learning technologies, modern computational analysis tools can now process structured data simultaneously from text, tables, and images \citep{wang2024news}. This has made it possible to accurately extract carbon emission data from environmental reports published by Chinese listed companies according to the ESG reporting standards.

\section{Data}\label{sec:Institutional background, database, variables, and sample}

\subsection{Carbon Disclosure Dataset}

\begin{figure}[htbp]
    \centering
    \includegraphics[width=0.5\linewidth]{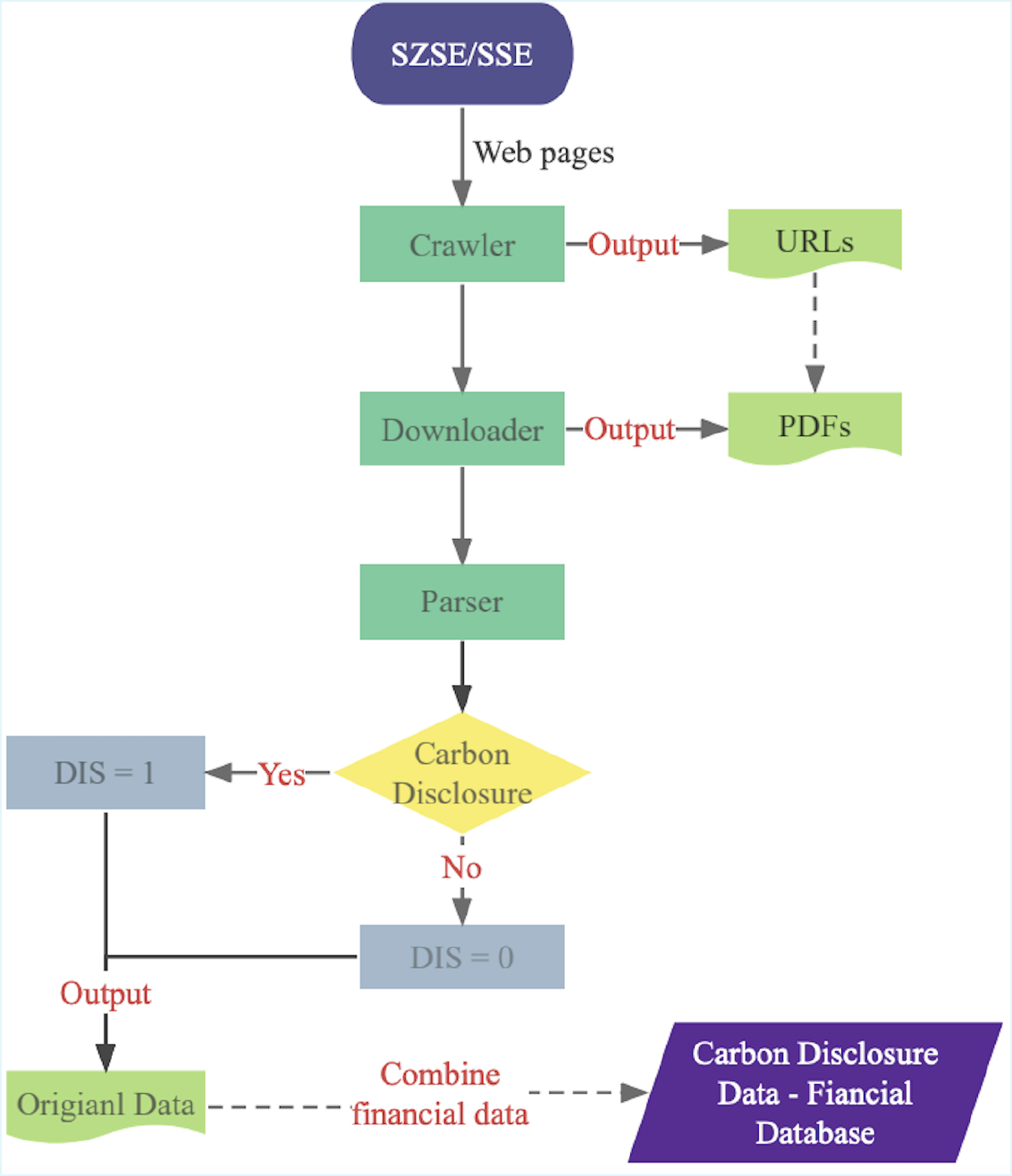}
    \caption{Flow chart of building carbon disclosure data-financial dataset}
    \label{fig:database flow chart}
\end{figure}
Given the limited availability of carbon emissions data for Chinese companies, with national databases only providing data until 2014-2015, we construct a novel dataset that captures carbon disclosure information up to June 2022. We define ``carbon disclosure'' as the presence of carbon-related information in corporate environmental reports, identified through specific keywords and accompanied by quantitative emissions data.

Our data construction process follows four main steps (see Fig. \ref{fig:database flow chart}). First, we use Python crawlers to collect environmental reports from the official websites of Shenzhen and Shanghai Stock Exchanges for the period 2018-2022. Second, we employ the pdfplumber library to extract carbon-related information, leveraging its capability to analyze machine-generated PDFs with structured forms. Third, we develop a rule-based algorithm that systematically identifies carbon disclosure based on predefined keywords (Table \ref{tab:keywords}). The algorithm examines both textual content and tabular data, specifically focusing on quantitative emissions information. Finally, we merge the extracted carbon disclosure data with financial information from Wind and CSMAR databases to create a comprehensive dataset.

This methodology enables us to efficiently process a large volume of corporate reports while maintaining accuracy in identifying carbon disclosures. The resulting dataset provides a unique opportunity to examine carbon disclosure practices across Chinese A-share listed companies in the context of evolving environmental regulations.

\begin{table*}[htbp]
\centering
\caption{Major Carbon Disclosure Information Keywords}
\scalebox{0.95}{
\begin{threeparttable}
\footnotesize
\begin{tabular}{p{3cm} p{10.5cm}} 
\toprule
\textbf{Meaning} & \textbf{Expressions} \\
\midrule
Emission of greenhouse gases & 
Total greenhouse gas emissions (Scopes 1, 2, and 3); total GHG emissions (Scopes 1 and 2); total GHG emissions; greenhouse gas emissions; CO$_2$ emissions; carbon emissions; GHG emissions; carbon footprint; climate emissions; etc. \\

Scope 1 emission & 
Direct greenhouse gas emissions (Scope 1); Scope 1 GHG emissions; Scope 1 CO$_2$ emissions; Scope 1: direct emissions; direct carbon emissions; etc. \\

Scope 2 emission & 
Indirect greenhouse gas emissions (Scope 2); Scope 2 GHG emissions; Scope 2 CO$_2$ emissions; Scope 2: indirect emissions; location-based Scope 2 emissions; energy-related carbon emissions; etc. \\
\bottomrule
\end{tabular}
\begin{tablenotes}
  \footnotesize
  \item  This table summarizes commonly used expressions for different categories of greenhouse gas emissions in carbon disclosure reports. Expressions are separated by semicolons and representative rather than exhaustive.
\end{tablenotes}
\end{threeparttable}
}
\label{tab:keywords}
\end{table*}

\subsection{Variables}

\subsubsection{Dependent Variables}

Our study investigates the relationship between carbon disclosure and financial performance. We focus on four dependent variables that capture different aspects of financial performance: rate of return (Return), stock volatility (Vol), ROE, and Tobin's Q.

Return of return on stock price is a key measure of the change in the value of a stock over a specific period. It reflects a company's stock performance and its impact on investor returns. It focuses on a business's past performance and its impact on investor returns.

Stock volatility measures stock price fluctuations over a given period. As a key indicator of stock risk, higher volatility is often associated with greater investment risk. This variable focuses on measuring potential risks involved in investing in a stock and the market's perception of the company's future performance.

ROE is a widely used ratio that measures a company's net income relative to its total equity. It is an important indicator of a company's profitability and reflects the level of compensation received by owners' equity. This variable focuses on a company's current profitability and its ability to generate returns for shareholders.

Tobin's Q measures a company's market value relative to its replacement cost. It is often used to gauge a company's overall financial performance and its ability to generate returns for shareholders. Generally, the higher Tobin's Q of a company, the greater its development potential. This variable focuses on a company's potential for future growth.

In conclusion, rate of return, stock volatility, ROE, and Tobin's Q are four key dependent variables in finance and economics research that provide valuable insights into various aspects of a company's financial performance.

\subsubsection{Control Variables}
To isolate the effect of carbon disclosure on financial performance, we include a comprehensive set of firm-level characteristics as control variables. These variables capture various aspects of firm operations, financial structure, and corporate characteristics that may influence both disclosure decisions and financial outcomes.

First, we control for firm size (Size), measured as the natural logarithm of total assets. According to \citet{hart1994strategy}, larger companies may possess different levels of resources and economies of scale that can impact their financial performance. The operational efficiency of firms is captured by two turnover ratios. Fixed asset turnover (FAT) measures how effectively a company utilizes its fixed assets to generate revenue, with higher ratios indicating more efficient asset utilization \citep{al2013effects}. Similarly, total assets turnover (TAT) reflects a company's overall efficiency in using its assets to generate revenue \citep{hapsoro2019does}.

Second, we include measures of financial structure and liquidity. The current ratio (CR) assesses a company's ability to meet short-term obligations, where a higher ratio indicates stronger liquidity position \citep{saleem2011impacts}. The asset-liability ratio (ALR) captures financial leverage, calculated as total liabilities divided by total assets. Additionally, we control for R\&D intensity using the natural logarithm of R\&D costs, as innovation capabilities may affect both environmental practices and financial outcomes.

Third, we incorporate several categorical variables to capture firm characteristics. Two dummy variables represent ownership structure: overseas listing status (Oversea) and state ownership (State-owned). Industry classification is captured by High-carbon, a dummy variable that equals one if a firm belongs to carbon-intensive sectors such as power generation, steel, cement, or chemicals. This classification is particularly relevant as high-carbon industries face different regulatory pressures and market expectations regarding carbon disclosure. These institutional features are especially pertinent in China's context, as they may influence both a firm's disclosure practices and its financial performance. All financial variables are winsorized at the 1st and 99th percentiles to mitigate the influence of outliers. The detailed variable definitions are provided in Table \ref{tab:variable definition}.

\subsubsection{Explanatory Variable}
Our study employs three categories of explanatory variables to capture different aspects of carbon disclosure and market participation. The first category focuses on carbon market participation mechanisms. \textit{$Treat^{National}$} is a dummy variable indicating whether a firm is included in the national carbon market, as designated by the National Ecological Environment Bureau. Similarly, \textit{$Treat^{local}$} identifies firms participating in local carbon markets, based on local ecological environment bureau records.

The second category measures carbon disclosure behavior. \textit{$Treat^{DIS}$} is a dummy variable that equals 1 if a firm discloses carbon emission information, determined through our analysis of environmental reports from the Shanghai and Shenzhen Stock Exchanges. This variable captures firms' voluntary disclosure practices and their compliance with environmental reporting requirements.

The third category comprises policy timing indicators that capture different implementation phases. \textit{$Post^{local}$} indicates the period after a firm's inclusion in local carbon markets, while \textit{$Post^{dis}$} marks the period following a firm's initial carbon information disclosure. Additionally, \textit{$Post^{dual}$} represents the period after the implementation of the dual-carbon target policy, capturing the broader policy environment shift.

The interaction of these treatment and timing indicators enables us to analyze how carbon disclosure and market participation affect firm performance across different policy regimes and market contexts. This comprehensive set of variables allows us to disentangle the effects of market mechanisms, disclosure behaviors, and policy implementations on corporate financial outcomes.

\begin{table}[H]
  \caption{The definition and data source of variables}
  \label{tab:variable definition}
  \centering
  \footnotesize
  \renewcommand{\arraystretch}{1}
  \setlength{\tabcolsep}{1mm}{
  \begin{tabular}{lp{0.6\columnwidth}p{0.2\columnwidth}}
    \hline
    Variable & Definition & Data Source \\
    \hline
    \textit{$Treat^{National}$} & Dummy variable for inclusion in national carbon market. If included in the national carbon market, its value is one; otherwise, its value is zero. & National ecological environment bureau \\
    \textit{$Treat^{local}$} & Dummy variable for inclusion in local carbon markets. If included in the local carbon market, its value is one; otherwise, its value is zero. & Local ecological environment bureau \\
    \textit{$Post^{local}$} & Dummy variable taking the value of one after being included in the local carbon market, zero otherwise. & Local ecological environment bureau \\
    \textit{$Treat^{DIS}$} & Dummy variable for carbon emission information disclosure. If a disclosure is made, its value is one; otherwise, zero. & Official website of SSE and SZSE \\
    \textit{$Post^{dis}$} & Dummy variable taking the value of one after carbon emission information disclosure, zero otherwise. & Official website of SSE and SZSE \\
    \textit{$Post^{dual}$} & Dummy variable taking the value of one after being included in the local carbon market, zero otherwise. & -- \\
    \textit{High-carbon} & Dummy variable for industry classification. If a company is in high-carbon industries, its value is one; otherwise, zero. & Wind \\
    \textit{Return} & Rate of return; the calculation period is weekly, calculated from the equal weighted average. & Wind \\
    \textit{Ln(Price)} & Log value of stock price; the calculation period is daily, calculated from the equal weighted average. & Wind \\
    \textit{Vol} & Volatility; the calculation period is weekly. & Wind \\
    \textit{VaR} & Value-at-risk; the calculation period is weekly, and the confidence level is 95\%. & Wind \\
    \textit{ROE} & Return on equity & Wind \\
    \textit{FAT} & Fixed asset turnover & Wind \\
    \textit{TAT} & Total assets turnover & Wind \\
    \textit{CR} & Current ratio & Wind \\
    \textit{ALR} & Asset liability ratio & Wind \\
    \textit{Size} & Log value of total asset; Total asset is measured in CNY. & Wind \\
    \textit{R\&D} & Log value of R\&D costs; R\&D costs are measured in CNY. & Wind \\
    \textit{Oversea} & Dummy variable for overseas listing. If listed overseas, its value is one; otherwise, zero. & Wind \\
    \textit{State-owned} & Dummy variable for whether it is a state-owned company. If so, value is one; otherwise, zero. & Wind \\
    \textit{Tobin's Q} & Market value divided by assets' replacement cost. & CSMAR \\
    \hline
  \end{tabular}}
\end{table}

\subsection{Sample and Descriptive Statistics}

Our sample consists of 2,298 Chinese A-share companies listed from 2018 to 2022, obtained from a combination of sources, including Wind Database, CSMAR, the Shanghai and Shenzhen Stock Exchanges and China National Knowledge Infrastructure (CNKI). To ensure data quality, companies that were not listed on the Shanghai and Shenzhen A-shares, underwent special treatment, or lacked necessary data for our analysis, as well as those in the finance and financial services industries, were excluded from the sample.

The industries of the sample companies were classified based on the system established by the China Securities Regulatory Commission (CSRC) and China's eight major industry sectors. This classification divided the sample into high-carbon and low-carbon industries. 
This classification system is considered to be more scientifically sound as it takes into account both industry characteristics and emission intensity.

Table \ref{tab:summary statistics} provides the descriptive statistics of the main variables included in our analysis.

\begin{table}[H]
  \caption{The statistical result of variables}
  \label{tab:summary statistics}
  \centering
  \footnotesize
  \begin{threeparttable}
  \setlength{\tabcolsep}{1mm}
  \renewcommand{\arraystretch}{1.2}
  \begin{tabular}{p{2cm}rrrrrrrr}
    \hline
    Variable & N & Mean & S.D. & Min & 25\% & Median & 75\% & Max \\
    \hline
    Nation carbon market & 90896 & 0.0054 & 0.0735 & 0 & 0 & 0 & 0 & 1 \\
    Local carbon market & 90896 & 0.0178 & 0.1321 & 0 & 0 & 0 & 0 & 1 \\
    Post local & 90896 & 0.0137 & 0.1164 & 0 & 0 & 0 & 0 & 1 \\
    DIS & 90896 & 0.0353 & 0.1846 & 0 & 0 & 0 & 0 & 1 \\
    Post dis & 90896 & 0.0161 & 0.1259 & 0 & 0 & 0 & 0 & 1 \\
    Post dual & 90896 & 0.3684 & 0.4824 & 0 & 0 & 0 & 1 & 1 \\
    High-carbon & 90896 & 0.7584 & 0.4281 & 0 & 1 & 1 & 1 & 1 \\
    Return & 71059 & 0.2262 & 4.2195 & -93.8819 & -0.1802 & -0.0023 & 0.2215 & 635.0785 \\
    Vol & 71106 & 6.5956 & 101.5892 & 0.0000 & 3.9222 & 5.3750 & 7.3604 & 27045.2240 \\
    VaR & 71846 & 7.5949 & 4.0044 & -61.0903 & 4.9696 & 7.1173 & 9.6755 & 74.5523 \\
    ROE & 77588 & 0.0272 & 1.8425 & -186.5570 & 0.0124 & 0.0404 & 0.0863 & 281.9892 \\
    FAT & 82904 & 14.9880 & 312.3227 & -19.2507 & 0.9216 & 2.1136 & 4.9291 & 55200.7410 \\
    TAT & 80258 & 0.4295 & 0.4429 & -0.1383 & 0.1527 & 0.3148 & 0.5673 & 11.9755 \\
    CR & 80255 & 2.7522 & 3.6187 & 0.0055 & 1.2301 & 1.7834 & 2.9840 & 235.4652 \\
    ALR & 80295 & 41.8940 & 98.3184 & 0.6171 & 24.5952 & 39.7723 & 55.2340 & 19173.8860 \\
    Size & 79821 & 22.0597 & 1.4365 & 15.3766 & 21.0668 & 21.9036 & 22.8694 & 28.6364 \\
    R\&D & 60027 & 16.4680 & 1.6233 & -4.6052 & 15.5880 & 16.4640 & 17.4084 & 23.6736 \\
    Oversea & 90896 & 0.0383 & 0.1918 & 0 & 0 & 0 & 0 & 1 \\
    State-owned & 90896 & 0.2184 & 0.4131 & 0 & 0 & 0 & 0 & 1 \\
    Tobin's Q & 70774 & 2.0323 & 2.0864 & 0.0162 & 1.2120 & 1.5570 & 2.1766 & 129.9253 \\
    \hline
  \end{tabular}
  \begin{tablenotes}
    \footnotesize
    \item This table presents the summary statistics of the variables in our main analysis for the mean, median, standard deviation (STD), minimum (Min), 25\% percentile (Q1), 75\% percentile (Q3), and maximum (Max) distributions. Variable definitions are detailed in Table \ref{tab:variable definition}.
  \end{tablenotes}
  \end{threeparttable}
\end{table}

\section{Empirical Models}\label{sec: Empirical results}

\subsection{Baseline Regression}

To assess the impact of carbon disclosure on corporate financial performance, we employ a fixed-effects difference-in-difference (DID) approach. Companies are categorized into two groups based on their carbon emission levels: high-carbon industries and low-carbon industries. Furthermore, in line with the implementation of the dual-carbon target, we divide the timeline into two periods: from January 1, 2018, to June 30, 2020, and from October 1, 2020, to December 31, 2021. This classification enables us to examine the effects of carbon disclosure on companies in different  industries both before and after the dual-carbon targets were implemented.

We use the DID approach to assess the impact of carbon disclosure on financial performance, using \textit{Return}, \textit{Vol}, \textit{ROE}, and \textit{Tobin's Q} as the dependent variable, the company's carbon disclosure is used as the independent variable. The model form is as follows:

\begin{equation}
Y_{i,t} = 
\begin{cases}
\alpha_{0} + \beta_{1} Treat^{1}_{i} \times post^{dis}_{t} + \gamma Z_{i,t} + \delta_{t} + \varphi_{i} + \varepsilon_{i,t}, & \text{if } t \in \text{2018Q1 to 2020Q2}, \\
\alpha_{0} + \beta_{1} Treat^{2}_{i} \times post^{dis}_{t} + \gamma Z_{i,t} + \delta_{t} + \varphi_{i} + \varepsilon_{i,t}, & \text{if } t \in \text{2020Q4 to 2021Q4}.
\end{cases}
\label{equ:adjusted_model}
\end{equation}

 where $Y_{i,t}$ represents \textit{Return}, \textit{Vol}, \textit{ROE} and \textit{Tobin's Q} for company $i$ at time $t$, where $t$ is the time of the company's first carbon disclosure, indexed in discrete quarterly terms, such as 2020Q1. Then, $t+1$ represents the subsequent quarter (e.g., if $t$ is 2020Q1, then $t+1$ is 2020Q2), and $t-1$ represents the previous quarter (e.g., if $t$ is 2020Q2, then $t-1$ is 2020Q1). $Treat^{1}_{i}$ ($Treat^{2}_{i}$) assumes a value of one if a company has made carbon disclosure between 2018Q1 and 2020Q2 (2020Q4 and 2021Q4); otherwise, its value is zero. $post^{dis}_{t}$ is a 
dummy variable equal to 1 after company i has made carbon disclosure; otherwise, its value is zero. i indexes the company, and t indexes the quarter. $Z_{i,t}$ denotes a vector of control variables (\textit{TAT}, \textit{FAT}, \textit{CR}, \textit{ALR}, \textit{Size}). 
$\delta_{t}$ represents the time-fixed effect,  $\varphi_{i}$ represents the entity-fixed effects, and $\varepsilon_{i,t} $ represents the random error item.

\begin{landscape}
\centering
\begin{table}[!t]
\scriptsize 
\renewcommand{\arraystretch}{0.8} 
\centering
\vspace{-20.0mm}
\begin{minipage}[!t]{\columnwidth}
  \renewcommand{\arraystretch}{1}
  \caption{Panel A: Effect of disclosure on high-carbon industries}
  \label{tab:disclosure}
  \centering

  \setlength{\tabcolsep}{1.5mm}{
\begin{tabular}{lllllllll}
\hline
                          & (1)         & (2)           & (3)           & (4)          & (5)         & (6)             & (7)          & (8)             \\\cmidrule(lr){2-5} \cmidrule(lr){6-9}
                          & \multicolumn{4}{l}{Before dual-carbon target: 2018/01/01 - 2020/06/30}  & \multicolumn{4}{l}{After dual-carbon target: 2020/10/01 - 2021/12/31}              \\ \cmidrule(lr){2-2} \cmidrule(lr){3-3} \cmidrule(lr){4-4} \cmidrule(lr){5-5} \cmidrule(lr){6-6} \cmidrule(lr){7-7} \cmidrule(lr){8-8} \cmidrule(lr){9-9} 
Variables                 & Return      & Vol            & ROE          & Tobin's Q   & Return          & Vol              & ROE              & Tobin's Q     \\ \hline
\textit{post $\times$  $treat_1$}      & -0.0295     & -0.0445           & -0.0129      & 0.076  &                 &              &                &                  \\
\textit{}                 & (-0.52)     & (-0.23)           & (-0.73)      & (0.70)     &                 &              &                &                    \\
\textit{post $\times$  $treat_2$}      &             &               &               &                 & \textbf{0.176}$^{*}$ & \textbf{-0.650}$^{*}$  & \textbf{0.0616}$^{*}$ & \textbf{0.357}$^{**}$     \\
\textit{}                 &             &               &               &            & (1.74)          & (-1.95)          & (1.91)           & (2.35)       \\
\textit{FAT}              & -0.00000139 & -0.0000580$^{***}$  & -0.00000979$^{*}$ & 0.00000261 & -0.0000457$^{**}$  & -0.000261$^{***}$ &  -0.0000492       & -0.0000637$^{***}$ \\
\textit{}                 & (-0.60)     & (-11.74)           & (-1.98)      & (0.30)      & (-3.21)         & (-13.28)         & (-1.10)          & (-7.49)       \\
\textit{TAT}              & 0.0813      & 0.367$^{**}$          & 0.187        & 0.130$^{***}$  & 0.0207          & 0.303            & 0.332$^{**}$          & 0.171$^{***}$        \\
\textit{}                 & (1.75)      & (3.23)            & (1.05)       & (2.78)      & (0.12)          & (1.57)           & (2.84)           & (3.51)        \\
\textit{CR}               & -0.0183$^{*}$     & 0.0269$^{*}$        & -0.0195$^{**}$     & -0.0149        & 0.0425          & 0.0521$^{*}$         & -0.0216          & -0.0254$^{**}$      \\
\textit{}                 & (2.13)      & (1.96)                & (-3.25)      & (-0.78)      & (1.21)          & (2.26)             & (-1.08)          & (-2.06)       \\
\textit{ALR}              & 0.000101    & -0.000365          & -0.0174$^{***}$   & 0.00607$^{*}$  & 0.0000413       & 0.0000360      & -0.0160          & 0.00604$^{***}$    \\
\textit{}                 & (1.26)      & (-0.35)              & (-3.35)      & (1.93)     & (0.95)          & (0.54)           & (-0.66)          & (255.69)      \\
\textit{Size} & -0.0630$^{*}$    & -0.622$^{***}$       & 0.295$^{**}$    & -1.952$^{*}$   & 0.239           & 0.699          & 0.689$^{*}$           & -0.953$^{***}$     \\
                          & (-2.55)     & (-13.64)           & (2.64)       & (-1.79)     & (0.91)          & (1.66)           & (2.24)           & (-6.52)       \\
Firm fixed effect         & Yes                  & Yes          & Yes         & Yes             & Yes          & Yes            & Yes              & Yes           \\
Quarter fixed effect         & Yes         & Yes           & Yes               & Yes             & Yes          & Yes            & Yes              & Yes           \\
Observations              & 25877       & 26120             & 28865        & 26143       & 16215           & 16035               & 17413            & 16039   \\ \hline     
\end{tabular}}
  \end{minipage}
  \\[12pt]
 \centering
    \begin{threeparttable}
  \begin{minipage}[!t]{\columnwidth}
  \renewcommand{\arraystretch}{1}
  Panel B: Effect of disclosure on low-carbon industries 
  \vspace{2.0mm}
  \centering
  \setlength{\tabcolsep}{3mm}{
\begin{tabular}{lllllllll}
\hline
                          & (1)         & (2)           & (3)           & (4)          & (5)         & (6)             & (7)          & (8)             \\\cmidrule(lr){2-5} \cmidrule(lr){6-9}
                          & \multicolumn{4}{l}{Before dual-carbon target: 2018/01/01 - 2020/06/30}  & \multicolumn{4}{l}{After dual-carbon target: 2020/10/01 - 2021/12/31}              \\ \cmidrule(lr){2-2} \cmidrule(lr){3-3} \cmidrule(lr){4-4} \cmidrule(lr){5-5} \cmidrule(lr){6-6} \cmidrule(lr){7-7} \cmidrule(lr){8-8} \cmidrule(lr){9-9} 
Variables                 & Return      & Vol            & ROE          & Tobin's Q   & Return          & Vol              & ROE              & Tobin's Q     \\ \hline
\textit{post $\times$  $treat_1$}      & 0.0289       & -0.737         & -0.0603     & \textbf{1.430}$^{**}$     &                &           &                &                \\
\textit{}                 & (0.22)       & (-1.20)          & (-0.83)     & (2.56)      &                &              &                 &           \\
\textit{post $\times$  $treat_2$}      &              &               &              &                  & {0.108} & -0.00761  & {0.0150} & 1.122  \\
\textit{}                  &               &              &             &             & (0.59)         & (-0.01)          & (0.25)          & (1.22)   \\
\textit{FAT}              & 0.000173     & -0.0000575     & -0.000133   & -0.000502$^{*}$   & 0.000484       & -0.000635   & -0.0000629      & -0.000175$^{*}$  \\
\textit{}                 & (1.85)       & (-0.28)            & (-1.38)     & (-1.81)     & (0.88)         & (-1.47)        & (-1.48)         & (-1.84)   \\
\textit{TAT}              & -0.0451      & -0.0593         & 0.0949      & 0.0032      & -0.363         & 0.0348          & 0.101$^{*}$          & -0.0523   \\
\textit{}                 & (-1.34)      & (-0.67)            & (1.24)      & (0.13)      & (-0.60)        & (0.19)          & (1.98)          & (-1.14)   \\
\textit{CR}               & 0.0340$^{**}$     & 0.0435$^{*}$           & -0.0125     & 0.00330     & -0.0125        & 0.0278         & -0.0338$^{***}$      & -0.0137   \\
\textit{}                 & (2.67)       & (2.48)           & (-0.46)     & (0.14)      & (-0.31)        & (1.19)       & (-3.43)         & (-1.09)   \\
\textit{ALR}              & -0.000962    & -0.000434        & -0.0243     & 0.00668$^{**}$   & -0.000651      & 0.000491        & -0.0267$^{***}$      & 0.0138$^{***}$ \\
\textit{}                 & (-1.37)      & (-0.17)         & (-1.65)     & (2.75)      & (-0.23)        & (0.12)        & (-4.36)         & (3.78)    \\
\textit{Size} & -0.146$^{**}$     & -0.627$^{***}$       & 0.487       & -1.416$^{**}$    & -0.823$^{*}$        & -0.0158           & 0.554$^{***}$        & -1.671$^{***}$ \\
                          & (-3.22)      & (-8.81)          & (1.91)      & (-3.34)     & (-2.35)        & (-0.04)     & (3.86)          & (-6.15)   \\
Firm fixed effect         & Yes          & Yes           & Yes          & Yes                 & Yes       & Yes            & Yes             & Yes       \\
Quarter fixed effect         & Yes          & Yes           & Yes                & Yes            & Yes       & Yes            & Yes             & Yes       \\
Observations              & 8635         & 8684                & 9451        & 8684        & 5189           & 5137           & 5547            & 5137      \\ \hline
\end{tabular}}
\end{minipage}
  \vspace{-0.1cm}
      \begin{tablenotes}
            \footnotesize
            \item This table reports the results of the DID analysis designed to test the impact of disclosure on high-carbon and low-carbon industries. Variable definitions are reported in 
    Appendix A. All regressions include both entity-fixed effects and time-fixed effects. The t-statistics reported in parentheses are based on standard errors clustered at the company level. $^{***}$, $^{**}$, and $^{*}$ indicate significance at the 1\%, 5\% and 10\% levels, respectively.
          \end{tablenotes}
      \end{threeparttable}
\end{table}
\end{landscape}

\textbf{Results}: The results of the regression analysis are presented in Table \ref{tab:disclosure}. Columns (1) and (5) use the rate of return to evaluate stock market performance, while columns (2) and (6) use volatility to assess the risk performance of the stock market. Columns (3) and (7) use ROE to gauge the fundamental performance, and Columns (4) and (8) use Tobin's Q to reflect the market's expectations for a company's future profit.

Our findings suggest that carbon disclosure has a significant impact on the financial performance of high-carbon industries after the implementation of the dual-carbon target. Prior to the implementation of the dual-carbon target, there was no significant change in financial performance for high-carbon industries. However, after the implementation of the dual-carbon target, the return rate of companies that disclosed carbon information increased significantly by 0.176, while the risk index ($Vol$) decreased significantly by 0.650. Additionally, ROE increased significantly by 0.0616, and Tobin's Q increased significantly by 0.357. These positive effects can be attributed to the reduction of the information gap between investors and corporate management, the enhancement of the company's ability to deal with future climate risks, and the increase in investor confidence resulting from carbon information disclosure.

For low-carbon industries, our results show almost no significant change in financial performance before and after the implementation of the dual-carbon target. The reason for the difference between high-carbon and low-carbon industries may be due to the greater vulnerability of high-carbon industries to climate risks in the context of dual-carbon target. Overall, our findings highlight the importance of carbon disclosure in improving the financial performance of high-carbon industries. Encourage carbon information disclosure to increase transparency and reduce information asymmetry between companies and investors to increase corporate climate change resilience.

\subsection{Dynamic Effects Model}
To alleviate our results are driven by reverse causation and examine the dynamic impact of carbon disclosure on corporate financial performance. We used the dynamic effects model. We included lagged dependent variables in the model as explanatory variables to reflect dynamic effects in our panel data analysis, an approach that has been widely adopted in studies examining the relationship between corporate environmental performance and financial performance \citep{elsayed2005impact}. Specifically, our dynamic effect test not only compares the difference between groups before the policy event, but also considers the differences between groups after the event. If the cross-items before period 0 (including period 0) are significant, it indicates that our model passes the parallel trend test. If the cross-items after period 0 (including period 0) are significant, the policy implementation has a lasting effect. The form of the model is as follows:
\begin{equation}
\begin{aligned}
Y_{i,t} =\ & \alpha_{0} + \beta_{1} Treat_{i} \times Before_{i,t}^{-4} + \beta_{2} Treat_{i} \times Before_{i,t}^{-3} + \beta_{3} Treat_{i} \times Before_{i,t}^{-2} \\
& + \beta_{4} Treat_{i} \times Before_{i,t}^{-1} + \beta_{5} Treat_{i} \times Current_{i,t}^{0} + \beta_{6} Treat_{i} \times After_{i,t}^{1} \\
& + \beta_{7} Treat_{i} \times After_{i,t}^{2} + \beta_{8} Treat_{i} \times After_{i,t}^{3} + \beta_{9} Treat_{i} \times After_{i,t}^{4} \\
& + \gamma Z_{i,t} + \delta_{t} + \varphi_{i} + \varepsilon_{i,t}
\end{aligned}
\label{equ:event_study_model}
\end{equation}

where $Before_{i,t}^{-4}$ ($Before_{i,t}^{-3}$, $Before_{i,t}^{-2}$ or $Before_{i,t}^{-1}$) is assigned a value of 1 if the observation is made four quarters (three quarters, two quarters, or one quarter) prior to carbon information disclosure and 0 otherwise. $Current_{i,t}^{0}$ is assigned a value of 1 if the observation is made in the quarter of carbon information disclosure and 0 otherwise. And $After_{i,t}^{-4}$ ($After_{i,t}^{-3}$, $After_{i,t}^{-2}$ or $After_{i,t}^{-1}$) is assigned a value of 1 if the observation is made four quarters (three quarters, two quarters, or one quarter) after carbon information disclosure and 0 otherwise. All other variables are the same as those described in the baseline DID regression.

\begin{figure}[h]
    \centering
    \includegraphics[width=\linewidth]{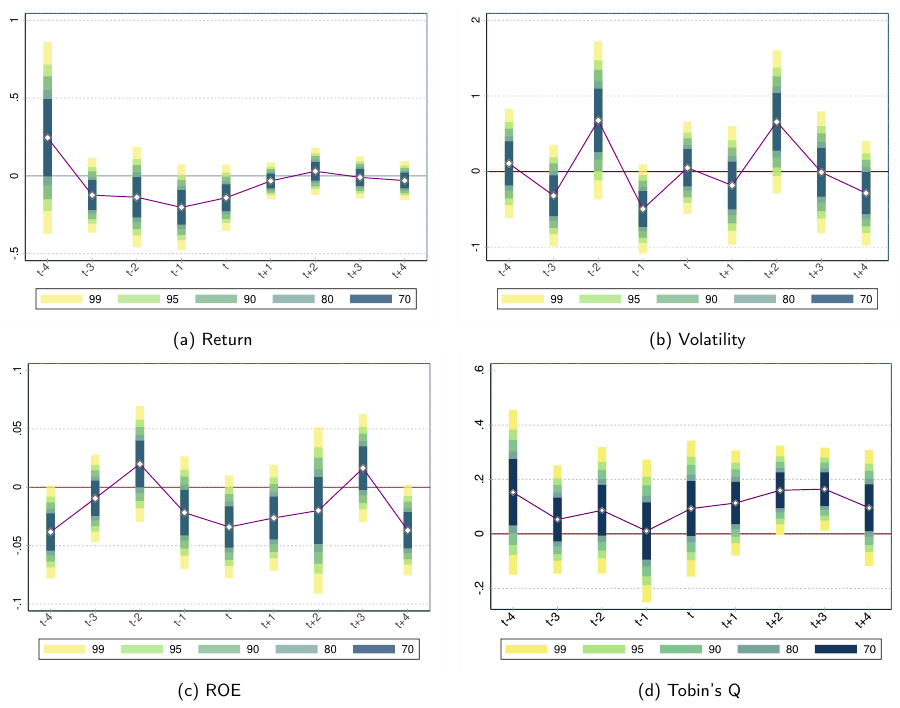}
    \caption{The dynamic effect of carbon information disclosure before the dual-carbon target}
    \label{fig:dynamic disclosure before}
\end{figure}

\begin{figure}[h]
    \centering
    \includegraphics[width=\linewidth]{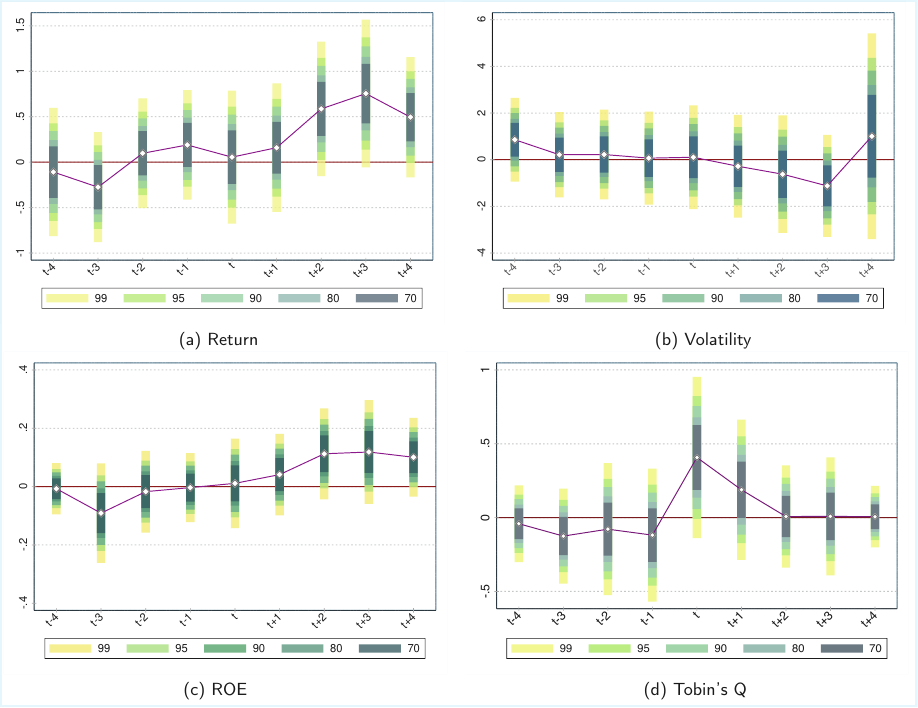}
    \caption{ The dynamic effect of carbon information disclosure after the dual-carbon target}
    \label{fig:dynamic disclosure after}
\end{figure}

Fig. \ref{fig:dynamic disclosure before} and Fig. \ref{fig:dynamic disclosure after} show the dynamic effects of carbon information disclosure before and after the dual-carbon target, respectively. To verify the parallel trends assumption, we studied the coefficients $\beta_1$, $\beta_2$, $\beta_3$, and $\beta_4$. Our findings indicate that, in most cases, the coefficients are not significant even at a 90\% confidence level, particularly after the dual-carbon target have been implemented, indicating that the parallel assumption of DID holds and alleviates the concern of reverse causation to some extent.

To illustrate the persistent impact of carbon disclosure on corporate financial performance, we focused on coefficients $\beta_5$, $\beta_6$, $\beta_7$, $\beta_8$, and $\beta_9$. We found that before the dual-carbon target, companies that disclosed carbon information had a persistent impact on Tobin's Q, but not on other dependent variables. This suggests that even before the dual-carbon target, the financial market was aware of the better development potential of companies that disclosed carbon information. 

\textbf{Results}: After the dual-carbon target, we found that both $Return$ and $ROE$ had a significant and persistent positive effect, while \textit{Tobin's Q} had a short-term positive effect but no obvious persistence. Additionally, $Vol$ showed a significant negative effect after a period of time. These results indicate that after the dual-carbon target, companies that made carbon disclosure had sustained positive advantages in the stock and financial markets, with increased development potential in the short term and decreased stock market risk. Our findings suggest that carbon disclosure can give companies a sustainable competitive advantage.

\subsection{Identification Issues and Robustness Checks}
This section presents a series of robustness checks on the impact of carbon information disclosure to ensure the robustness of the empirical results. Placebo tests were employed to assess the randomness of the outcomes by randomly assigning a ``pseudo-treatment'' to the control group and re-estimating the model to validate the actual treatment effect. This was followed by applying Propensity Score Matching (PSM) to reduce selection bias, and matching the treatment and control groups on observed covariates to align their pre-intervention characteristics. Furthermore, the Two-Stage Least Squares (2SLS) model was utilized to address potential endogeneity issues, with instrumental variables predicting the values of endogenous explanatory variables in the first stage and using these predicted values to estimate causal effects in the second stage. The integration of these methodologies provided comprehensive robustness checks against potential omitted variable bias and endogeneity issues, thereby reinforcing the reliability of the research hypothesis.
\subsubsection{Placebo Tests}
To avoid the possibility of our DID analysis results being accidental, we performed a placebo test. The traditional placebo test randomly selects the same number of companies as the experimental groups as the ``pseudo-treatment group'' and the remaining companies as the control group, then generates ``pseudo-policy dummy variables'' for regression. This method is suitable for the classic DID model where policy times are consistent, but it is not applicable in our case of time-varying DID where each unit has a different policy time.

Therefore, we first recorded the disclosure times of all the original experimental group samples and then randomly assigned these disclosure times to other companies. The selected companies became the experimental group, and the disclosure time was their selected time. By randomly assigning the treatment and control groups, we expected the impact of the ``pseudo-policy dummy variables'' on stock prices to be zero, otherwise, the policy effects we obtained in the previous section would be unreliable \citep{ijerph18168839}. Additionally, if carbon disclosure does indeed have a positive impact on stock prices, we would expect the truly estimated coefficients to be greater than the placebo effect.

We repeated the process 500 times, and the results are presented in Table \ref{tab: placebo}. The table summarizes the distribution of the coefficients of $Treat \times Post$ from the DID regressions by reporting the mean, 5th percentile, 25th percentile, median, 75th percentile, and 95th percentile. Our findings show that the estimated coefficients are mostly centered around zero, and most p-values are greater than 0.05 (not significant at the 5\% level). This suggests that our results are unlikely to have been obtained by chance and are robust, unaffected by other policies or random factors.

\begin{table}[H]
\caption{Placebo test}
  \begin{threeparttable}
\setlength{\tabcolsep}{0.5mm}{
\begin{tabular}{p{1.3cm}lllllllll}
\hline
Variable & Actual             & Mean    & p5                & p10              & p25     & p50     & p75             & p90            & p95     \\ \hline
Return   & $\textbf{0.176}^{*}$    & -0.0206 & -1.023            & -0.5266          & -0.1913 & -0.018  & 0.1461 & 0.4501         & 1.0157  \\
Vol      & \textbf{-0.650}$^{*}$  & 0.0068  & -0.7703$^{*}$           & -0.5691 & -0.282  & -0.0078 & 0.2875          & 0.5803         & 0.7974$^{*}$  \\
ROE      & \textbf{0.0616}$^{*}$    & -0.0126 & -0.1943           & -0.0717          & -0.0207 & -0.0012 & 0.0158          & 0.047 & 0.0953  \\
Tobin's Q   & \textbf{0.252}$^{*}$  & -0.0017   & -0.1606 & -0.1174          & -0.0600 & -0.0047  & 0.0574          & 0.1216         & 0.1594 \\ \hline
\end{tabular}}
\begin{tablenotes}
        \footnotesize
        \item  $^{***}$, $^{**}$, and $^{*}$ indicate significance at the 1\%, 5\% and 10\% levels, respectively.
      \end{tablenotes}
  \end{threeparttable}
\label{tab: placebo}
\end{table}

\subsubsection{Propensity Score Matching}
 To estimate the treatment effect with reduced selection bias and endogeneity, we employ the Propensity Score Matching-Difference-in-Differences (PSM-DID) method. This technique uses observational data to analyze the impact of interventions and is rooted in logistic regression. The propensity score represents the probability of a sample being part of the treatment group, and by matching these scores, PSM helps ensure that the treatment and control groups are similar in terms of observed characteristics.

In this study, we enhance the robustness of our results by applying PSM in conjunction with the DID model. The equation used in our analysis is given below. We use the expectation operator (denoted by E(·)), and the parameters $\Delta API^{1}_{i,t}$ and $\Delta API^{0}_{i,t}$ represent the relative financial performance change in the treatment and control groups, respectively. The parameter $E(\Delta API^{0}_{i,t}\mid DIS_{i}=1)$ is a ``counterfactual'' that indicates the change in financial performance if the company did not make a carbon disclosure.

\begin{equation}
    ATT_{i,t} = E(\Delta API^{1}_{i,t}\mid DIS_{i}=1)-E(\Delta API^{0}_{i,t}\mid DIS_{i}=1)
\end{equation}

\begin{table}[htbp]
\centering
\caption{The Pseudo $R^{2}$ and Joint Statistical Significance of the Covariate Propensity Score}
\setlength{\tabcolsep}{8.8mm}{
\begin{tabular}{llll}
\hline
    Type & Pseudo $R^{2}$ & LR chi2 & P\textgreater chi2 \\ \hline
  Unmatched     & 0.271                                & 1444.96            & 0.000                        \\
             Matched       & 0.000                                & 0.52             & 0.992     \\ \hline
\end{tabular}}
\label{tab:pseudo}
\end{table}

\begin{table}[htb]
\caption{Balance tests of PSM (high carbon industry)}
\setlength{\tabcolsep}{1.4mm}{
\begin{tabular}{llllllll}
\hline
Covariate & Type &\multicolumn{2}{c}{Mean}      & Bias(\%) & \multirow{2}{*}{\makecell{Reduct\\Bias(\%)}} & \multicolumn{2}{c}{T-test}                    \\ \cline{3-4} \cline{7-8} 
\textbf{}          & \textbf{}     & Treated &Control & \textbf{}         & \textbf{}                & t   & P \textgreater $\left| t \right|$ \\ \hline
\textit{FAT}                & Unmatched     & 3.4732            & 10.092            & -3.3              &                          & -0.57            & 0.566                       \\
                   & Matched       & 3.3152            & 3.2258            & 0.0             & 98.6                     & 0.30           & 0.767                       \\
\textit{TAT}                & Unmatched     & 0.5281            & 0.4649            & 15.9              &                          & 4.08            & 0.000                       \\
                   & Matched       & 0.5236            & 0.5225            & 0.3               & 98.2                     & 0.04            & 0.968                       \\
\textit{CR}                & Unmatched     & 1.7103           & 2.9096           & -42.6             &                          & -8.02           & 0.000                       \\
                   & Matched       & 1.7471           & 1.7442           & 0.1              & 99.8                     & 0.02           & 0.980                       \\
\textit{ALR}                & Unmatched     & 50.742            & 39.354            & 60.9             &                          & 13.99           & 0.000                       \\
                   & Matched       & 50.07            & 50.576            & -2.7              & 95.6                     & -0.46           & 0.643                       \\
\textit{Size}                & Unmatched     & 24.467           & 22.061           & 165.0              &                          & 44.08            & 0.000                       \\
                   & Matched       & 24.256           & 24.256           & -0.0              & 100.0                     & -0.00           & 1.000                      \\\hline
\end{tabular}}
\label{tab:bias psm}
\end{table}

To match the samples, we used the radius matching method (with a radius of 0.001). After matching, we checked the balance of the distribution of covariates between the experimental and control groups, and the results are shown in Table \ref{tab:bias psm}. We also report the pseudo $R^{2}$ and joint statistical significance of the covariate propensity score in Table \ref{tab:pseudo}. As can be seen from Table \ref{tab:bias psm}, many variables had significant differences between the treatment and control groups before PSM matching, but they all became insignificant after PSM matching. The pseudo $R^{2}$ of the matched samples is close to 0 and its P-value becomes insignificant, as shown in Table \ref{tab:pseudo}. This suggests that the difference in economic performance between the treatment and control groups can only be attributed to the carbon disclosure after PSM matching. Hence, the DID model combined with the PSM method effectively estimates the causal effect of a specific policy on relevant outcomes.

We present the PSM-DID results in Table \ref{tab: psm}. The coefficient significance of the cross term $post \times treat$ does not change, indicating that our estimation results are robust.

\begin{table}[h]
\caption{Propensity score matching}
\setlength{\tabcolsep}{1mm}{
\begin{tabular}{ccccc}
\hline
                     & (1)          & (2)          & (3)        & (4)           \\
Variables            & Return       & Vol          & ROE        & Tobin's Q     \\ \hline
\textit{post $\times$  $treat$}             & \textbf{0.179}$^{*}$         & \textbf{-0.661}$^{*}$        & \textbf{0.0578}$^{*}$      & \textbf{0.368}$^{**}$         \\
\textit{}                 & (1.71)       & (-1.87)      & (1.80)     & (2.36)        \\
\textit{FAT}              & -0.0000420$^{***}$ & -0.000276$^{***}$ & -0.0000492 & -0.0000643$^{***}$ \\
                          & (-2.82)      & (-12.89)     & (-1.10)    & (-7.50)       \\
\textit{TAT}              & 0.0202       & 0.288        & 0.333$^{***}$    & 0.160$^{***}$      \\
\textit{}                 & (0.11)       & (1.50)       & (2.83)     & (3.35)        \\
\textit{CR}               & 0.0479       & 0.0319       & -0.0216    & -0.0264$^{**}$      \\
\textit{}                 & (1.34)       & (1.62)       & (-1.08)    & (-2.06)       \\
\textit{ALR}              & 0.00875      & -0.0330$^{**}$     & -0.0160    & 0.00468       \\
                          & (1.13)       & (-2.29)      & (-0.66)    & (1.58)        \\
\textit{Size} & 0.207        & 1.160$^{**}$       & 0.689$^{**}$     & -0.936$^{***}$     \\
\textit{}                 & (0.85)       & (2.17)       & (2.24)     & (-5.52)       \\
Firm fixed effect      & Yes          & Yes          & Yes        & Yes           \\
Quarter fixed effect         & Yes          & Yes          & Yes        & Yes           \\
Obsevations               & 16085        & 15908        & 17375      & 15922         \\ \hline
\end{tabular}}
\label{tab: psm}
\end{table}

\subsubsection{Two-stage Least-squares Model}
To alleviate concerns about endogeneity driven by simultaneity and reverse causality, we employ an instrumental variable (IV) approach to examine the impact of carbon information disclosure on corporate financial performance. We carefully selected two instrumental variables -- lagged dependent variable and the female director ratio. Based on previous research, the lagged dependent variable is often chosen as an instrumental variable in papers on information disclosure \citep{buallay2020sustainability,cui2018does}. The female director ratio refers to the percentage of women on a company's board of directors. Previous research has found that female directors in China have a positive impact on corporate carbon disclosure \citep{he2021female}. However, there is no direct influence of this ratio on corporate financial performance \citep{rose2007does}. Based on the above, we employ a 2SLS as follows.

First-stage: 
\begin{equation}
CID_{i,t} = \beta_{0} + \beta_{1}CID_{i,t-1}+\beta_{2}FEMALE\_RATIO_{i,t} + \sum_{j = 3}^{n} \beta_j Z_{i,t} + \delta_{t} + \varphi_{i} + \varepsilon_{i,t}    
\end{equation}

Second-stage: 
\begin{equation}
Y_{i,t} = \gamma_{0} + \gamma_{1}\widehat{CID_{i,t}} + \sum_{j = 2}^{n} \gamma_j Z_{i,t} + \delta_{t} + \varphi_{i} + \varepsilon_{i,t}    
\end{equation}

where $CID_{i,t}$ is carbon information disclosure, which is $post \times treat$ in baseline regression. $CID_{i,t-1}$ is the one-period lagged $CID_{i,t}$. $\widehat{CID}_{i,t}$ is the estimated value from the first stage. $FEMALE\_RATIO_{i,t}$ is female director ratio. All other variables are the same as those described in the baseline DID regression.

The results of this analysis, which can be seen in Table \ref{tab: 2sls}, are consistent with the baseline results and provide further evidence to mitigate concerns about reverse causality. The coefficient significance of the cross term $CID_{i,t}$ ($post \times treat$) remains unchanged, providing further confidence in the robustness of our results.

\begin{table}[h]
\caption{Two-stage least-squares model}
\setlength{\tabcolsep}{1mm}{
\begin{tabular}{ccccc}
\hline
                     & (1)          & (2)          & (3)        & (4)           \\
Variables            & Return       & Vol          & ROE        & Tobin's Q     \\ \hline
\textit{post $\times$  $treat$}             & \textbf{0.176}$^{*}$       & \textbf{-0.650}$^{**}$      & \textbf{0.0616}$^{**}$    & \textbf{0.357}$^{***}$    \\
\textit{}                 & (1.78)      & (-2.29)      & (2.18)     & (2.92)     \\
\textit{FAT}              & -0.0000457$^{**}$ & -0.000261$^{***}$ & -0.0000492 & -0.0000637$^{*}$ \\
\textit{}                 & (-2.44)     & (-3.70)      & (-1.05)    & (-1.83)    \\
\textit{TAT}              & 0.0207      & 0.303        & 0.332$^{***}$    & 0.171$^{***}$   \\
\textit{}                 & (0.12)      & (1.50)       & (2.65)     & (3.79)     \\
\textit{CR}               & 0.0425      & 0.0521$^{***}$     & -0.0216    & -0.0254$^{***}$  \\
\textit{}                 & (1.41)      & (2.61)       & (-1.33)    & (-2.74)    \\
\textit{ALR}              & 0.0000413   & 0.0000360    & -0.0160    & 0.00604$^{***}$ \\
                          & (0.90)      & (0.16)       & (-0.80)    & (17.93)    \\
\textit{Size} & 0.239       & 0.699$^{*}$        & 0.689$^{**}$     & -0.953$^{***}$  \\
                          & (0.96)      & (1.91)       & (2.49)     & (-7.99)    \\
Firm fixed effect      & Yes         & Yes          & Yes        & Yes        \\
Quarter fixed effect      & Yes         & Yes          & Yes        & Yes        \\
Observations              & 16132       & 15942        & 17394      & 15955             \\ \hline
\end{tabular}}
\label{tab: 2sls}
\end{table}

\subsection{Determinants and Motivations of Carbon Disclosure}\label{sec:Determinants of carbon disclosure}

\subsubsection{Determinants of Carbon Disclosure}

In this section, we employ a logistic regression model to examine the determinants of corporate carbon disclosure. The analysis focuses on six factors: R\&D intensity (\textit{R\&D}), overseas listing status (\textit{Oversea}), carbon market participation (\textit{Local carbon market}), state ownership (\textit{State-owned}), high‑carbon industry affiliation (\textit{High-carbon}), and the dual‑carbon target period (\textit{$Post^{dual}$}).

The regression results (Table \ref{tab: disclosure determinant}) show that all six factors have a positive and statistically significant effect on the likelihood of carbon information disclosure. These findings indicate that both external pressures and self-competitive advantages play a role in prompting companies to disclose their carbon information.

\begin{table}[H]
 \centering
 \caption{Determinants of corporate carbon disclosure}
 \setlength{\tabcolsep}{8mm}{

\begin{tabular}{ll}
\hline
                                        & $Post^{dis}$ \\ \hline
\textit{R\&D}                           & 0.107$^{***}$     \\
\textit{}                               & (4.16)       \\
\textit{Oversea}                        & 1.164$^{***}$     \\
\textit{}                               & (10.23)      \\
\textit{Carbon market}            & 0.599$^{***}$    \\
\textit{}                               & (3.59)       \\
\textit{State-owned}                    & 0.193$^{*}$       \\
\textit{}                               & (2.36)       \\
\textit{High-carbon}                    & 0.862$^{*}$       \\
\textit{}                               & (2.09)       \\
\textit{$Post^{dual}$} & 0.469$^{*}$        \\
\textit{}                               & (2.42)       \\
\textit{FAT}                            & -0.00377     \\
\textit{}                               & (-1.88)      \\
\textit{TAT}                            & 0.477$^{***}$    \\
\textit{}                               & (5.33)       \\
\textit{CR}                             & -0.0481      \\
\textit{}                               & (-1.75)      \\
\textit{ALR}                            & -0.0106$^{***}$  \\
\textit{}                               & (-3.92)      \\
\textit{Size}                & 0.931$^{***}$     \\
\textit{}                               & (26.56)      \\
Firm fixed effect                   & Yes          \\
Quarter fixed effect                    & Yes          \\
Observations                            & 44972        \\ \hline
\label{tab: disclosure determinant}
\end{tabular}}
\end{table}

\subsubsection{Carbon Disclosure from an Environmental Score Perspective}

To determine whether companies in high-carbon and low-carbon industries choose to disclose carbon information due to external pressure or self-assurance, we analyzed the distribution of environmental scores among companies with and without carbon disclosure. As a relevant environmental indicator, carbon emissions data is not yet available in China's A-share market. Therefore, we used the company's environmental score as a proxy to reflect its carbon score.
To uncover the possible motives behind carbon disclosure, we obtained environmental score data from the Wind database. The scores were normalized to ensure comparability across different years and were divided into two groups based on whether or not the companies disclosed carbon information in each year from 2018 to 2021.

\begin{figure}[htbp]
    \centering
    \includegraphics[width=0.75\linewidth]{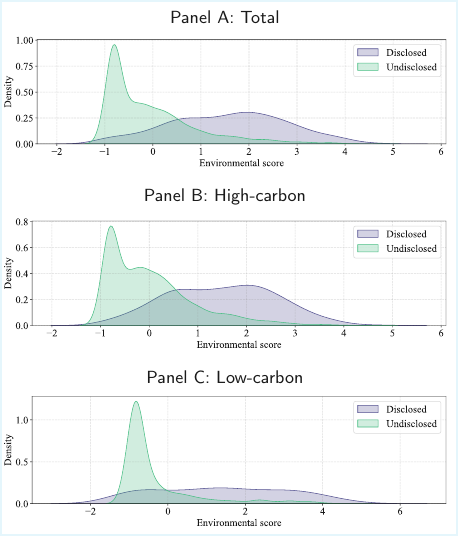}
    \caption{The corporate environmental score for whether or not to disclose carbon information}
    \label{fig: E-score distribution}
\end{figure}

The environmental score distribution is presented in Fig. \ref{fig: E-score distribution}. The results indicate that companies that choose to disclose carbon information have significantly higher environmental scores, with a clear trend observed in low-carbon industries. This may suggest that low-carbon industries have higher levels of self-confidence, as external pressure plays a lesser role in their decision-making. In high-carbon industries, however, external pressure may play a more significant role, leading to a higher number of companies with poor environmental performance still disclosing carbon information.

In conclusion, our findings support that companies' self-assurance in their carbon emissions data can act as a competitive advantage and motivate them to disclose this information.

\section{Conclusion}\label{sec:Conclusion and policy implications}
This study assesses the impact of carbon disclosure on the financial performance of companies listed on China's A-share market and explores the determinants in corporate carbon disclosure behavior following the implementation of the dual-carbon targets. By constructing a carbon disclosure-finance data set that includes 4,336 companies, this research reveals that, after the implementation of the dual carbon target, carbon disclosure significantly improves financial performance in high-carbon industries, whereas no significant effects are observed in low-carbon industries. Furthermore, factors such as R\&D, overseas listing status, participation in the national/regional carbon market, state-owned status, belonging to a high-carbon industry, and the post-dual-carbon target period significantly improve the probability of corporate carbon disclosure. 
Based on these findings, it is recommended that policymakers and corporate managers recognize the strategic importance of carbon disclosure and actively promote and refine relevant regulatory frameworks while further exploring how to optimize carbon disclosure through technological and policy innovations. In summary, this study highlights the critical role of enhanced carbon disclosure in addressing global climate change challenges, providing robust theoretical and empirical support to improve environmental transparency and maintain competitive advantage in financial markets.

\bibliography{ref}
\end{document}